\documentclass{tMPH2e}

\usepackage{amsmath,amsfonts,%amsthm,
amssymb}

\newcommand{\version}{October 1, 2015}
\newcommand{\upchi}{\raise1pt\hbox{$\chi$}}

\newcommand{\cE}{{\mathord{\cal E}}}

\begin{document}
\title{A New Look at Thomas-Fermi Theory} %
\author{Jan Philip Solovej\thanks{This work was supported by the European Research Council under
ERC Advanced Grant project. no. 321029}\\
    \normalsize Department of Mathematics, 
    \normalsize University of Copenhagen,
        \normalsize DK-2100 Copenhagen, Denmark\\
        \normalsize {\it e-mail\/}: solovej@math.ku.dk\\
  \bigskip
{\it Dedicated to Andreas Savin}
  \date{\version}\bigskip
}

                                \maketitle

\begin{abstract}
  In this short note we argue that Thomas-Fermi Theory the simplest of
  all density functional theories, although failing to explain
  features such as binding or stability of negative ions, is
  surprisingly accurate in estimating sizes of atoms.  We give both
  numerical, experimental and rigorous mathematical evidence for this
  claim. Motivated by this we formulate two new mathematical
  conjectures on the exactness of Thomas-Fermi Theory.
\end{abstract}

\vfill\eject
\section{Introduction}

With the enormous success of density functional theories in
computational chemistry and the refinement they have undergone in the
past half century (see \cite{burke} for a recent review of successes
and challenges in DFT) it may be surprising if there should still be
anything to be learned from as old and basic a model as Thomas-Fermi
(TF) Theory (Thomas 1927 \cite{thomas1927} and Fermi 1927
\cite{fermi1927}).  TF Theory goes back to the early days of quantum
mechanics and is, indeed, the oldest density functional theory.

Thomas-Fermi Theory completely ignores exchange correlation effects
and is unable to predict many basic properties of atoms and
molecules. Atoms do not bind in Thomas-Fermi Theory and negatively
charged ions are unstable (Teller's No-Binding Theorem
\cite{teller1962}).
 
It is nevertheless the purpose of this short note to show that TF
Theory, surprisingly enough, is quite adequate at describing certain
specific and at the same time important properties of atoms and
molecules. As an example we shall see that the size of alkali atoms is
not just qualitatively, but even from a quantitative point of view
rather accurately described by TF Theory. We shall also briefly
discuss evidence that TF theory accurately describes the short
distance behavior of the Born-Oppenheimer energy curves of diatomic
molecules. As a consequence of the No-Binding Theorem there is no
equilibrium point in the TF approximation to the Born-Oppenheimer
curve, but at much shorter distances the TF curve may as we shall
discuss give a surprisingly good approximation.  

We will discuss the new point of view on the validity of TF theory
both in a stringent mathematical formulation but also based on
experimental and numerical evidence.  The paper is organized as
follows. In Section~\ref{sec:largeZ} we briefly review Thomas-Fermi
Theory and give its mathematical justification based on the large $Z$
asymptotics of the ground state energy.  In
Section~\ref{sec:newasymptotics} we discuss new asymptotic limits
which may be accurately described by TF theory and which we believe
are closer at capturing real and important chemical aspects of atoms
and molecules than the ground state energy asymptotics.
Finally, in Section~\ref{sec:numerics} we discuss
simple numerical calculations of the radii of atoms in TF Theory which, at least
for the larger alkali atoms give extraordinarily good agreement with
the empirical data of Bragg \cite{bragg1920} and Slater
\cite{slater1964} (see Table~\ref{tab:1} below).

\section{Thomas-Fermi Theory and the  large $Z$ limit}\label{sec:largeZ}
Thomas-Fermi Theory for a molecule or atom is defined from the energy functional (in atomic units)
$$
\cE^{\rm TF}(\rho) =\frac3{10}(3\pi^2)^{2/3}\int \rho(r)^{5/3}
-V(r)\rho(r) d^3r+\frac12\iint\frac{\rho(r)\rho(r')}{|r-r'|}d^3rd^3r'+U,
$$
where $V$ is the electron-nuclear attraction and $U$ is the internuclear repulsion, i.e.,
$$
V(r)=\sum_{k=1}^M\frac{Z_k}{|r-R_k|},\quad U=\sum_{1\leq k<\ell\leq M}\frac{Z_kZ_\ell}{|R_k-R_\ell|}.
$$
Here we have $M$ nuclei at positions $R_k$, $k=1,2,\ldots,M$. In particular, for an atom 
$$U=0\quad\text{and}\quad  
V(r)=\frac{Z}{|r|}.
$$
It is not difficult to see that for an atom the minimizing energy satisfies
$$
E^{\rm TF}(Z)=\min_\rho \cE^{\rm TF}(\rho)  =-C_{\rm TF} Z^{7/3}
$$
for some constant $C_{\rm TF}>0$ whose numerical value does not play a role to our
discussion.  The minimizing density $\rho^{\rm TF}$ satisfies
$\int\rho^{\rm TF}=Z$, i.e., the atom is neutral and solves the
TF-equation
$$
\frac12(3\pi^2)^{2/3} \rho^{\rm TF}(r)^{2/3}=V(r)-\int\frac{\rho^{\rm TF}(r')}{|r-r'|}d^3r'.
$$ 
If we minimize with the restriction that $\int\rho(r')d^3r'=N<Z$ we
arrive at the Thomas-Fermi energy $E^{\rm TF}(N,Z)$ of a positive
ion. As already pointed out negative ions are unstable.

The mathematical justification of Thomas-Fermi Theory was given in
\cite{liebsimon1973,liebsimon1977} (see also \cite{lieb1981}). In
these works TF Theory is justified as giving the correct leading order ground state
energy asymptotics of atoms for large atomic number $Z$. The limit $Z$
tending to infinity is of course in some sense purely academic. It is,
however, necessary, in order to state mathematically that an
approximation becomes exact, to consider a precise limiting situation,
even if that takes us out of the physically interesting regime.  The
large $Z$ limit is in general also a good test for approximating
schemes, see e.g.\ \cite{CSPB}.

It is known that the large $Z$ asymptotics of the exact 
total ground state energy of an atom is  given by 
$$
E(Z)=-C_{\rm TF}Z^{7/3}+C_{\rm S}Z^2+C_{\rm DS}Z^{5/3}+o(Z^{5/3}),
$$
where the leading term is, indeed, the Thomas-Fermi energy as
established in \cite{liebsimon1973}.  We are using the convention that
a quantity, such as the total ground state energy $E(Z)$, without a
superscript, denotes the exact value whereas the approximation in, say,
Thomas-Fermi Theory is denoted with a superscript $E^{\rm TF}(Z)$. 

In atomic units $C_{\rm S}=1/2$ and this $Z^2$ correction was
predicted by Scott in \cite{scott1952} and proved in
\cite{hughes1986,siedentopweikard1987}. Finally, the last term of
order $Z^{5/3}$ was derived by Schwinger \cite{schwinger1981} based
partly on Dirac's exchange estimate \cite{dirac1930} and established
mathematically in a series of papers by Fefferman and Seco (see
\cite{feffermanseco}).

To the order $Z^{5/3}$ the asymptotics above agrees \cite{bach,grafsolovej}
with the ground state energy in Hartree-Fock Theory 
as well as with the ground state energy in Kohn-Sham theory 
with the exchange-correlation energy given by the Dirac exchange term 
$$
-C_{\rm D}\int\rho(r)^{4/3}d^3r.
$$

Establishing the ground state energy to order $Z^{5/3}$ is a
mathematical tour de force and has led to a wealth of beautiful
mathematics. Unfortunately, this order is far from what is needed in
order to control quantities of interest to chemistry such as
ionization energies, atomic radii, molecular bond lengths, or Born-Oppenheimer
energy curves. These quantities are all expected to be of order $O(1)$
in the large $Z$ limit.  This claim is known in the mathematical
physics literature as the {\it ionization conjecture}
\cite{simonproblems,liebseiringer,bachbook}.
Establishing the accuracy of any approximating scheme for the
total ground state energy to the order $O(1)$ is beyond reach of any known
method.  In particular, improving the expansion above 
does not seem to be a promising route to prove the ionization
conjecture and the conjecture is, indeed, still
one of the outstanding problems in mathematical physics \cite{simonproblems}.

This does not rule out, however, that we may be able to establish the
adequate accuracy of approximating schemes in estimating ionization
energies, bond lengths, etc.  To illustrate this very simple point
recall that the first ionization energy is
$$
I_1(Z)=E(N=Z-1,Z)-E(Z),
$$
i.e., a difference of two quantities of order $Z^{7/3}$. Estimating
the total energies independently is of course a bad idea, when we are
interested in their difference.  In the next section we will propose
an asymptotic formula directly for the ionization energies and atomic radii. 

\section{New Asymptotic Limits}\label{sec:newasymptotics}
Our first observation is that the ionization conjecture, in fact, holds in Thomas-Fermi Theory. 
Much more is actually true. If we define the $m$-th ionization energy of an atom by
$$
I_m(Z)=E(N=Z-m,Z)-E(Z)
$$ 
then in Thomas-Fermi Theory we have that 
$$
\lim_{Z\to\infty}I^{\rm TF}_m(Z)= a^{\rm TF}m^{7/3}
$$
for some positive explicit constant $a^{\rm TF}\geq0$ which is not the same as $C^{\rm TF}$ above. 

Likewise, if we define the radius $R_m(Z)$ to the last $m$ electrons in an atom by
$$
\int_{|r|>R_m(Z)}\rho(r)d^3r=m
$$  
we have in Thomas-Fermi Theory
$$
\lim_{Z\to\infty}R^{\rm TF}_m(Z)=b^{\rm TF}m^{-3}
$$
for an explicit constant $b^{\rm TF}>0$ (in fact, $b^{\rm
  TF}=(81\pi^2/2)^{1/3}$).

In analogy to the the energy expansion we conjecture that the exact
values of $\lim_{Z\to\infty}I_m(Z)$ and $\lim_{Z\to\infty}R_m(Z)$ have
leading order asymptotic expansions given by the TF expressions above
when $m\to\infty$.  To be a little more precise we, in fact, do not
believe that the limits of $I_m(Z)$ and $R_m(Z)$ necessarily exist as
$Z\to\infty$. They may oscillate between upper and lower limits
(liminf and limsup). I.e., for the ionization energy we have the upper
and lower limits
$\underline{\lim}_{Z\to\infty}I_m(Z)\leq\overline{\lim}_{Z\to\infty}I_m(Z)$
and likewise for the radius $\underline{\lim}_{Z\to\infty}R_m(Z)\leq
\overline{\lim}_{Z\to\infty}R_m(Z)$. We conjecture that for both the
ionization energy and the radius the upper and lower limits have the
same asymptotics for large $m$ and that they are given by the
Thomas-Fermi expressions.  We are hence led to the following
conjecture.

{\bf Generalized Ionization Conjecture for Atoms:}
  {\it The large $Z$ (upper and lower) limits of the ionization energies
  and the radii of atoms satisfy the asymptotic formulas
\begin{eqnarray}
\mathop{\overline{\underline{\lim}}}\limits_{Z\to\infty}I_m(Z)&=&a^{\rm TF}m^{7/3}+o(m^{7/3})\label{eq:Imconj}\\
\mathop{\overline{\underline{\lim}}}\limits_{Z\to\infty}R_m(Z)&=&b^{\rm TF}m^{-1/3}+o(m^{-1/3})\label{eq:Rmconj}
\end{eqnarray}
as $m\to\infty$.
}

A strong theoretical argument in favor of this conjecture is that it
was proved to hold in Hartree-fock Theory in \cite{solovej2003}. It
should be emphasized that when we talk about Hartree-Fock Theory we
mean the totally unrestricted theory where the minimization is over
all Slater determinants, i.e., not restricted to any particular basis
set. Whereas the proof in the case of Hartree-Fock Theory does not
generalize to the full many-body quantum context it should apply to a
variety of models, e.g., some restricted Hartree-Fock models or
Kohn-Sham density functional theories with local exchange correlation
terms.\footnote{I would like to thank Andreas Savin for pointing out
  that there could be an interest in generalizing the result beyond
  Hartree-Fock Theory} 
In Section~\ref{sec:numerics} below we present some experimental
evidence that the approximation in (\ref{eq:Rmconj}) may not only be
asymptotically good, but even remarkably accurate for real atoms.

We turn next to the generalized ionization conjecture for molecules,
which is concerned with the asymptotics of the Born-Oppenheimer curves
for diatomic molecules. For simplicity let us consider neutral
homonuclear diatomic molecules with two nuclei of charge $Z$ situated
a distance $R$ from each other.  We denote the ground state energy by
$E_{\rm mol}(Z,R)$ and subtract the dissociation energy $2E(Z)$.
As a consequence of the No-Binding Theorem we have in Thomas-Fermi
Theory that
$$
E^{\rm TF}(Z,R)-2E^{\rm TF}(Z)>0.
$$
Moreover, this expression has a universal behavior \cite{brezislieb}
$$
\lim_{Z\to\infty}\left(E^{\rm TF}(Z,R)-2E^{\rm TF}(Z)\right)=D^{\rm TF}R^{-7}. 
$$
This leads to the following molecular version of the ionization conjecture. 

{\bf Generalized Ionization Conjecture for Molecules:}
{\it The large $Z$  (upper and lower) limits of the homonuclear diatomic Born-Oppenheimer curve satisfies the 
asymptotic formula
\begin{equation}\label{eq:BOconj}
\mathop{\overline{\underline{\lim}}}\limits_{Z\to\infty}
\left(E(Z,R)-2E(Z)\right)=D^{\rm TF}R^{-7}+o(R^{-7}).
\end{equation}
as $R\to0$. 
}

This conjecture is still open even in Hartree-Fock Theory, but I
strongly believe that it can be settled in that case. 
Strong numerical evidence that the TF Born-Oppenheimer curve is a good approximation at short internuclear 
distances was presented in \cite{gilkasolovejtaylor}.

\section{Comparison of $R^{\rm TF}_m(Z)$ with empirical radii}\label{sec:numerics}
\begin{figure}
\begin{center}
\includegraphics[width=0.8\hsize]{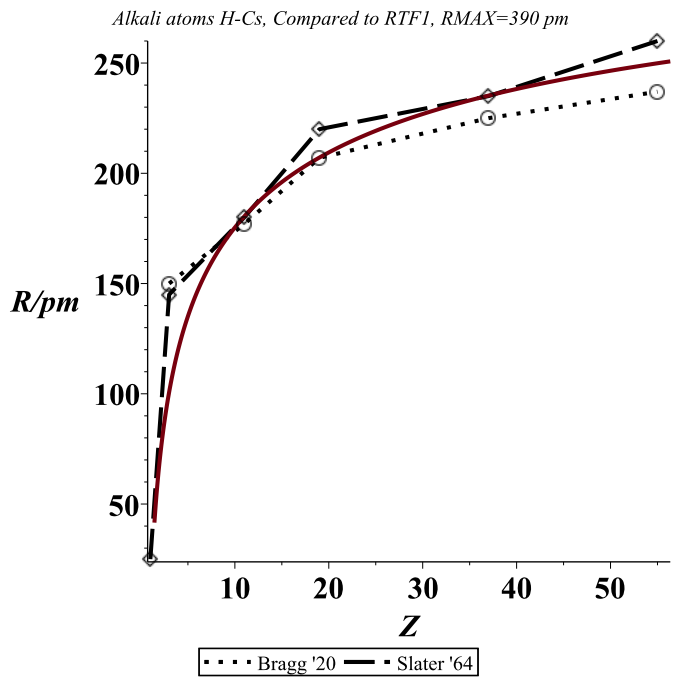}
\vspace{-4cm}
\caption{\label{fig1} The Thomas-Fermi radius $R_1$ plotted against
the Bragg and Slater empirical radii for alkali elements}
\end{center}
\end{figure}

Here we finally present the results of comparing numerical
calculations of the Thomas-Fermi radii $R^{\rm TF}_m(Z)$ with the
empirical radii found in \cite{bragg1920} and \cite{slater1964}.  It is
natural to assume that the approximation is best for the case of
alkali atoms with one electron outside a closed shell atom, i.e., that
the radius $R_1(Z)$ is a good measure for the empirical
radius. Table~\ref{tab:1} shows the remarkably good comparison of the calculated values
of $R^{\rm TF}_1(Z)$ and the empirical values of Bragg and Slater.
\begin{table}%
\tbl{\label{tab:1} Radii of alkali atoms in pm. Empirical values of Bragg~\cite{bragg1920} 
and Slater~\cite{slater1964} compared to the values of
the Thomas-Fermi Radius $R_1$}
{\begin{tabular}{|l|c|c|c|}
\hline
Element&Bragg 1920&Slater 1964&TF radius $R_1$\\
\hline
Li&150&145&101\\
\hline
Na&177&180&181\\
\hline
Rb&225&235&235\\
\hline
K&207&220&207\\
\hline
Cs&237&260&250\\
\hline
Fr&?&?&266\\
\hline
\end{tabular}}
\end{table}
We have also considered the case of group 2 atoms. Here it is not
clear for which $m$, $R_m(Z)$ would be a good measure of the empirical
radius. It is natural to believe that it should be the same $m$
through the group and that $1<m<2$. Table~\ref{tab:2} shows the
comparison of the empirical radii with the calculated values for
$R^{\rm TF}(Z)$ for $m=1.4$.  Figure~\ref{fig1} Shows the graph of the
radius function $R_1^{\rm TF}(Z)$ again with empirical values of
Bragg and Slater.

\begin{table}
\tbl{\label{tab:2} Group 2 atomic radii in pm. Empirical values of Bragg~\cite{bragg1920} 
and Slater~\cite{slater1964} compared to the 
the Thomas-Fermi Radius $R_{1.4}$}
{\begin{tabular}{|l|c|c|c|}
\hline
Element&Bragg 1920&Slater 1964&TF radius $R_{1.4}$\\
\hline
Be&115&105&87\\
\hline
Mg&150&142&149\\
\hline
Ca&170&180&173\\
\hline
Sr&195&200&199\\
\hline
Ba&210&215&213\\
\hline
\end{tabular}
}
\end{table} 

To emphasize the simplicity in finding $R^{\rm TF}_m(Z)$ we
briefly comment on the calculation. If we introduce the Thomas-Fermi
potential for a neutral atom
$$
\phi^{\rm TF}(r)=\frac{Z}{|r|}-\int \frac{\rho^{\rm TF}(r')}{|r-r'|}d^3r'
$$
we have that 
$$
\int_{|r|>R}\rho^{\rm TF}(r)d^3r=-R^2\frac{d}{dr}\phi^{\rm TF}(R).
$$
Thus the defining equation for $R^{\rm TF}_m(Z)$ is 
$
-R^{\rm TF}_m(Z)^2\frac{d}{dr}\phi^{\rm TF}(R^{\rm TF}_m(Z))=m.
$
To determine $\phi^{\rm TF}$ we note that from the scaling property of Thomas-Fermi Theory 
$$\phi^{\rm  TF}(r)=2^{7/3}(3\pi)^{-2/3}Z^{4/3}\phi_0(Z^{1/3}2^{7/3}(3\pi)^{-2/3}r)
$$ 
for a positive function 
$\phi_0$ independent of $Z$ satisfying the differential equation
$$
-\Delta\phi_0(r)=\phi_0(r)^{3/2}, \quad r\ne0
$$
with $\lim_{r\to0}r\phi_0(r)=1$ and vanishing at infinity. 
Since $\phi_0$ is spherical it is straightforward to solve this as an ODE. 
We find that $\phi_0(r)=r^{-1}-1.588+o(1)$ for small $r$ and that
$\phi_0(r)=144r^{-4}(1-13.26r^{-\zeta})+o(r^{-4-\zeta})$ for large $r$, 
where $\zeta=\frac12(\sqrt{73}-7)$. 

Having found $\phi_0$ we have $R^{\rm
  TF}(Z)=2^{7/3}(3\pi)^{-2/3}Z^{-1/3} r$ where $r$ is determined from
$-r^2\phi_0'(r)=m/Z $.

\section{Conclusion}
We have proposed a new way to view the validity of the Thomas-Fermi
approximation which is more directly linked to properties of chemical
interest than the traditional point of view based on the total binding
energy.  We have given arguments for this new approach based both on
exact mathematical results for the Hartree-Fock model and comparisons
with experimental and numerical evidence.  In particular, we have
found surprisingly good (and probably better than expected) agreement
between the Thomas-Fermi calculated radii of atoms and empirical
values.
 
In analogy the ground state energy, where the correction terms to the TF
asymptotics are known, it would be very interesting if it would be
possible to find the next corrections to the proposed formulas
(\ref{eq:Imconj}), (\ref{eq:Rmconj}), and (\ref{eq:BOconj}).

\end{document}